\documentclass[aps,prc,twocolumn,superscriptaddress,showpacs,showkeys]{revtex4-1}

\usepackage{amsmath}
\usepackage{graphicx}
\usepackage{wasysym}

\usepackage{subfigure}
\usepackage{scrextend}
\usepackage[breaklinks]{hyperref}

\begin{document}


\title{First determination of $\beta$-delayed multiple neutron emission beyond $A=100$ through direct neutron measurement: The P$_{2n}$ value of $^{136}$Sb.}


	\author{R.~Caballero-Folch}
	\email[Author contact: ]{rcaballero-folch@triumf.ca/roger@baeturia.com}
	\affiliation{TRIUMF, Vancouver, British Columbia V6T 2A3, Canada}
	\author{I.~Dillmann}
	\affiliation{TRIUMF, Vancouver, British Columbia V6T 2A3, Canada}
	\affiliation{GSI Helmholtz Center for Heavy Ion Research, D-64291 Darmstadt, Germany}
	\author{J.~Agramunt} 
	\affiliation{IFIC, CSIC - Universitat de Val\`encia, E-46071 Val\`encia, Spain}
	\author{J.L.~Ta\'in} 
	\affiliation{IFIC, CSIC - Universitat de Val\`encia, E-46071 Val\`encia, Spain}
	\author{A.~Algora} 
	\affiliation{IFIC, CSIC - Universitat de Val\`encia, E-46071 Val\`encia, Spain}
	\affiliation{Institute of Nuclear Research of the Hungarian Academy of Sciences, Debrecen H-4001, Hungary}
	\author{J.~\"Ayst\"o}
	\affiliation{University of Jyv\"askyl\"a, FI-40014 Jyv\"askyl\"a, Finland}
	\affiliation{Helsinki Institute of Physics, University of Helsinki, FI-00014 Helsinki, Finland}
	\author{F.~Calvi\~no} 
	\affiliation{Universitat Polit\`ecnica de Catalunya, E-08028 Barcelona, Spain}
	\author{L.~Canete} 
	\affiliation{University of Jyv\"askyl\"a, FI-40014 Jyv\"askyl\"a, Finland}
	\author{G.~Cort\`es} 
	\affiliation{Universitat Polit\`ecnica de Catalunya, E-08028 Barcelona, Spain}
	\author{C.~Domingo-Pardo}
	\affiliation{IFIC, CSIC - Universitat de Val\`encia, E-46071 Val\`encia, Spain}
	\author{T.~Eronen} 
	\affiliation{University of Jyv\"askyl\"a, FI-40014 Jyv\"askyl\"a, Finland}
	\author{E.~Ganioglu}
	\affiliation{University of Istanbul, 34134 Vezneciler, Turkey}
	\author{W.~Gelletly}
	\affiliation{University of Surrey, Guildford GU2 7XH, United Kingdom}
	\author{D.~Gorelov} 
	\affiliation{University of Jyv\"askyl\"a, FI-40014 Jyv\"askyl\"a, Finland}
	\author{V.~Guadilla} 
	\affiliation{IFIC, CSIC - Universitat de Val\`encia, E-46071 Val\`encia, Spain}
	\author{J.~Hakala} 
	\affiliation{University of Jyv\"askyl\"a, FI-40014 Jyv\"askyl\"a, Finland}
	\author{A.~Jokinen} 
	\affiliation{University of Jyv\"askyl\"a, FI-40014 Jyv\"askyl\"a, Finland}
	\author{A.~Kankainen}
	\affiliation{University of Jyv\"askyl\"a, FI-40014 Jyv\"askyl\"a, Finland}
	\author{V.~Kolhinen}
	\affiliation{University of Jyv\"askyl\"a, FI-40014 Jyv\"askyl\"a, Finland}
	\author{J.~Koponen} 
	\affiliation{University of Jyv\"askyl\"a, FI-40014 Jyv\"askyl\"a, Finland}
	\author{M.~Marta}
	\affiliation{GSI Helmholtz Center for Heavy Ion Research, D-64291 Darmstadt, Germany}
	\author{E.~Mendoza} 
	\affiliation{CIEMAT, E-28040 Madrid, Spain}
	\author{A.~Montaner-Piz\'a}
	\affiliation{IFIC, CSIC - Universitat de Val\`encia, E-46071 Val\`encia, Spain}
	\author{I.~Moore}
	\affiliation{University of Jyv\"askyl\"a, FI-40014 Jyv\"askyl\"a, Finland}
	\author{C.R.~Nobs}
	\affiliation{University of Brighton, BN2 4AT Brighton, United Kingdom}
	\author{S.E.A.~Orrigo}
	\affiliation{IFIC, CSIC - Universitat de Val\`encia, E-46071 Val\`encia, Spain}
	\author{H.~Penttil\"a} 
	\affiliation{University of Jyv\"askyl\"a, FI-40014 Jyv\"askyl\"a, Finland}
	\author{I.~Pohjalainen} 
	\affiliation{University of Jyv\"askyl\"a, FI-40014 Jyv\"askyl\"a, Finland}
	\author{J.~Reinikainen} 
	\affiliation{University of Jyv\"askyl\"a, FI-40014 Jyv\"askyl\"a, Finland}
	\author{A.~Riego}
	\affiliation{Universitat Polit\`ecnica de Catalunya, E-08028 Barcelona, Spain}
	\author{S.~Rinta-Antila} 
	\affiliation{University of Jyv\"askyl\"a, FI-40014 Jyv\"askyl\"a, Finland}
	\author{B.~Rubio}
	\affiliation{IFIC, CSIC - Universitat de Val\`encia, E-46071 Val\`encia, Spain}
	\author{P.~Salvador-Casti\~neira}
	\affiliation{Universitat Polit\`ecnica de Catalunya, E-08028 Barcelona, Spain}
	\author{V.~Simutkin}
	\affiliation{University of Jyv\"askyl\"a, FI-40014 Jyv\"askyl\"a, Finland}
	\author{A.~Tarife\~no-Saldivia}
	\affiliation{IFIC, CSIC - Universitat de Val\`encia, E-46071 Val\`encia, Spain}
	\affiliation{Universitat Polit\`ecnica de Catalunya, E-08028 Barcelona, Spain}
	\author{A.~Tolosa-Delgado}
	\affiliation{IFIC, CSIC - Universitat de Val\`encia, E-46071 Val\`encia, Spain}
	\author{A.~Voss}
	\affiliation{University of Jyv\"askyl\"a, FI-40014 Jyv\"askyl\"a, Finland}


\date{\today}

\begin{abstract}
\noindent
\textbf{Background:} $\beta$-delayed multiple neutron emission has been observed for some nuclei with A$\leq$100, being the $^{100}$Rb the heaviest $\beta$2n emitter measured to date. So far, only 25 P$_{2n}$ values have been determined for the $\approx$300 nuclei that may decay in this way. Accordingly, it is of interest to measure P$_{2n}$ values for the other possible multiple neutron emitters throughout the chart of the nuclides. It is of particular interest to make such measurement for nuclei with A$>$100 to test the predictions of theoretical models and simulation tools for the decays of heavy nuclei in the region of very neutron-rich nuclei. In addition, the decay properties of these nuclei are fundamental for the understanding of astrophysical nucleosynthesis processes such as the $r$-process, and safety inputs for nuclear reactors.\\
\textbf{Purpose:} To determine for the first time the two neutron branching ratio, P$_{2n}$ value, for $^{136}$Sb through a direct neutron measurement, and to provide precise P$_{1n}$ values for $^{136}$Sb and $^{136}$Te.\\
\textbf{Method:} A pure beam of each isotope of interest was provided by the JYFLTRAP Penning trap at the Ion Guide Isotope Separator On-Line (IGISOL) facility of the University of Jyv\"askyl\"a, Finland. The purified ions were implanted into a moving tape at the end of the beam line. The detection setup consisted of a plastic scintillator placed right behind the implantation point after the tape to register the $\beta$-decays, and the BELEN detector, based on neutron counters embedded in a polyethylene matrix. The analysis was based on the study of the $\beta$- and neutron- growth-and-decay curves and the $\beta$-one-neutron and $\beta$-two-neutron time correlations, which allowed us the determination of the neutron-branching ratios.\\
\textbf{Results:} The P$_{2n}$ value of $^{136}$Sb was found to be 0.14(3)\% and the measured P$_{1n}$ values for $^{136}$Sb and $^{136}$Te were found to be 32.2(15)\% and 1.47(6)\%, respectively.\\
\textbf{Conclusions:} The measured P$_{2n}$ value is a factor 44 smaller than predicted by the finite-range droplet model plus the quasiparticle random-phase approximation (FRDM+QRPA) model used for $r$-process calculations.\\
\end{abstract}


\maketitle


\section{Introduction}\label{sec:intro}
The $\beta$ decay Q value (Q$_{\beta}$ value) increases when going towards neutron-rich nuclei in the chart of nuclides. This makes $\beta^{-}$ decay the dominant decay mode for this region. In very neutron-rich nuclei, the emission of one or more neutrons may also occur after a $\beta$-decay. This process of $\beta$-delayed neutron ($\beta n$) emission was discovered in 1939 by Roberts et al.~\cite{roberts1939further} and is energetically allowed when the Q$_{\beta n}$ value is positive, i.e. the Q$_{\beta}$ value of the decay exceeds the neutron separation energy (S$_{n}$) of the daughter nucleus. This phenomenon becomes dominant when the populated state in the daughter nucleus, following the $\beta$-decay, is higher in excitation energy than S$_{n}$, which gets lower going towards neutron-rich nuclei in the chart of nuclides. When the populated states lie even higher than the two-neutron separation energy, S$_{2n}$, i.e. Q$_{\beta 2n} > 0$, two neutron emission may also occur. This phenomenon was predicted in 1960 by Goldansky~\cite{goldansky1960neutron}. The first studies of multiple-neutron emission were carried out in the 1980s in which detection of two-neutron emission was observed for nuclei up to mass A=33 and predictions for the emission of three neutrons were made for masses around A=100~\cite{lyutostansky1985beta}. The $\beta$-delayed two-neutron ($\beta2n$) emission probability, P$_{2n}$ value, has only been experimentally determined for 25 isotopes (two of them only approximations)~\cite{audi2017nubase2016} of out of $\approx$300 potential multiple neutron emitter candidates~\cite{wang2017ame2016}. The importance of providing more precise data for neutron emitting isotopes has been highlighted by the IAEA Coordinated Research Projects in Refs.~\cite{IAEA2017INDC} and on \textit{$\beta$-delayed neutron emission evaluation}~\cite{IAEA2011INDC}. The latter emphasizes the importance of these data for safety improvement purposes for emerging nuclear power reactors, as well as for astrophysical studies. Indeed, these experimental data, involving isotopes in the neutron-rich region, are needed to achieve a better knowledge of the ``freeze-out" of the rapid neutron capture ($r$-process)~\cite{burbidge1957b2fh} when theoretical calculations of nucleosynthesis~\cite{arcones2011dynamical,mumpower2016impact,surman2014sensitivity} are applied. The delayed neutron emission is an important input for these models as it shifts the final isobaric solar system abundances of some species to lower masses, and introduces neutrons in the stellar environment that can be re-captured by other nuclei at later stages. This is of special interest in the regions of the $r$-abundance peaks such as A$\approx$130 near the doubly-magic $^{132}$Sn isotope (N=82 and Z=50). Up to now, the P$_{n}$ data available for the heavy mass region is scarce, especially above $A=150$~\cite{Caballero-Folch2016First}, and non-existent for multiple neutron emitters above $A=100$.\\

This paper presents a measurement of the P$_{2n}$ neutron branching ratio for $^{136}$Sb. With a Q$_{\beta 2n}$ window of 1884(6)~keV~\cite{wang2017ame2016}, $^{136}$Sb has been suggested to be a multiple-neutron emitter in several theoretical models~\cite{moller2003new,borzov2016self,mumpower2016neutron} and by some experiments: these include estimates based on the Finite Fermi-system theory~\cite{lyutostansky1983estimation} and more recently an experiment involving several isotopes of mass $A=136$ in which the isobars could not be separated well enough to determine their neutron branching ratios precisely~\cite{TESTOV-EspRus2011}. In order to get a successful measurement of multiple $\beta$-delayed neutron emission, we used a high efficiency neutron detector and a beam free from contamination of other isotopes. The pure $^{136}$Sb beam was obtained with the Penning trap mass spectrometer, JYFLTRAP, at the Ion Guide Isotope Separator On-Line (IGISOL) facility~\cite{aysto2001development,hakala2012precision}, located in the Accelerator Laboratory of the University of Jyv\"askyl\"a, Finland (JYFL). This contribution describes the experimental setup in Section~\ref{sec:setup}, the details of the data analysis following the methodology reported in~\cite{agramunt2016characterization} and the results in Sec.~\ref{sec:analysis}. A summary and discussion are presented in Sec.~\ref{sec:discussion}, and the conclusions are given in Sec.~\ref{sec:conclusions}.

\section{Experimental setup}\label{sec:setup}
The experiment presented in this paper was carried out at the IGISOL facility of JYFL. The isotopes of interest were produced in nuclear fission induced with a 25~MeV proton beam, with an intensity of around 7-10~$\mu A$, impinging on a natural uranium target. The fission fragments produced were extracted from the helium gas cell using a sextupole ion guide (SPIG)~\cite{karvonen2008sextupole} and differential pumping, accelerated to 30~keV and mass-separated with a dipole magnet. The continuous beam was cooled and bunched using an RFQ cooler-buncher device~\cite{nieminen2001time} before injecting the ions into the JYFLTRAP double Penning trap. The purification took place in the first trap, where a mass-selective buffer gas cooling technique~\cite{savard1991new} was employed to resolve different isobars based on their cyclotron resonance frequencies $\nu_{c}=qB/(2\pi m)$, where q and m are the charge and the mass of the ion of interest and B is the magnetic field strength inside the trap. The description and the principles of the JYFLTRAP at IGISOL are reported in Ref.~\cite{eronen2012jyfltrap}, and Figs. 1 and 2 of Ref.~\cite{kolhinen2013recommissioning} detail the layout of the IGISOL facility.\\

\begin{figure}[!ht]
\centering
\includegraphics[width = 0.8\columnwidth]{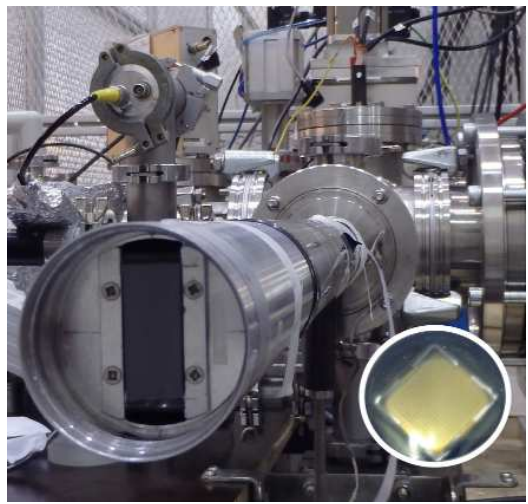}
\caption{(Color online) Aluminum tube linking JYFLTRAP and the end of the beamline. The rear side of the moving tape at the implantation position can be seen. The plastic scintillator detector used as a $\beta$ decay counter, shown in the bottom-right, was placed 6~mm behind the moving implantation tape at the end of the aluminum tube.}
\label{Fig:Tube}
\end{figure}

\begin{figure}[!ht]
\centering
\includegraphics[width = 0.8\columnwidth]{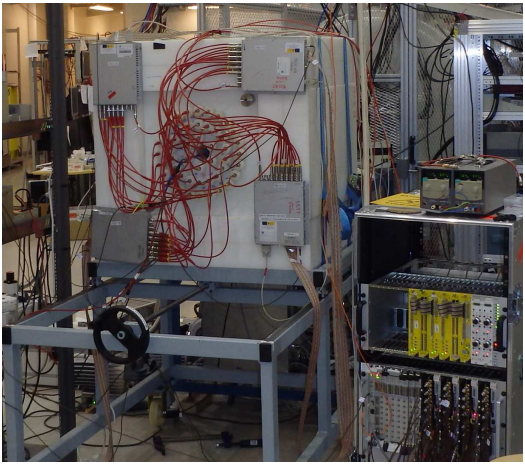}
\caption{(Color online) The BELEN detector embedded in the high-density polyethylene (HDPE) matrix and shielding, and the data acquisition system.}
\label{Fig:BELENSetup}
\end{figure}

In this experiment, pure ion beams of $^{95}$Rb, $^{137}$I, $^{136}$Te and $^{136}$Sb were extracted from the trap without isobaric contaminants and transported to the implantation system. $^{95}$Rb and $^{137}$I were used for calibration purposes, as their P$_{1n}$ values are well known from previous experiments~\cite{rudstam1993delayed,pfeiffer2002status,iaea2018evaluation}. The implantation system consisted of a moving magnetic tape placed inside an aluminum tube, with a thickness of 1~mm and diameter of 46~mm, under vacuum. This tube linked JYFLTRAP and the end of the beamline, shown in Fig.~\ref{Fig:Tube}.
 The moving tape system allowed control of the ion-implantation (beam on) and decay (beam off) times according to the half-life of the isotope measured, in order to be able to reproduce the growth-and-decay curves in the analysis (see Section~\ref{sec:analysis}). 
 The detection system in this experiment consisted of a 3~mm-thick plastic scintillator $\beta$ counter, shown in Fig.~\ref{Fig:Tube} (bottom-right), placed at the end of the vacuum tube, surrounded by the Beta dELayEd Neutron (BELEN) detector~\cite{TDRBELEN,BELEN-conf2010,BELEN-hyperfine}. The latter consisted of 48 $^3$He counter tubes of 2.54~cm diameter manufactured by LND Inc.~\cite{LNDInc}, distributed in three concentric rings, and embedded in a high-density polyethylene (HDPE) matrix to moderate the neutrons, see Fig.~\ref{Fig:BELENSetup}. BELEN was surrounded by 20~cm of HDPE shielding in order to moderate and absorb neutrons scattered from the surroundings (see also Fig.~\ref{Fig:BELENSetup}).
Table~\ref{table:BELEN-Charac} summarizes its characteristics.\\
\begin{table}[ht!]
\caption{Characteristics of the BELEN detector (units in millimiters)\label{table:BELEN-Charac}}
\begin{ruledtabular}
\begin{tabular}{ccccc}
&Central&\multicolumn{3}{c}{Ring}\\
&hole&Inner&Middle&Outer\\
\hline
\\
Position (diameter)& 60& 120& 230& 340\\
Number of counters (10 atm)& & 0& 8& 0\\
Number of counters (8 atm)& & 6& 10& 24\\
\end{tabular}
\end{ruledtabular}
\end{table}

\begin{figure}[!ht]
\centering
\includegraphics[width = 1.0\columnwidth]{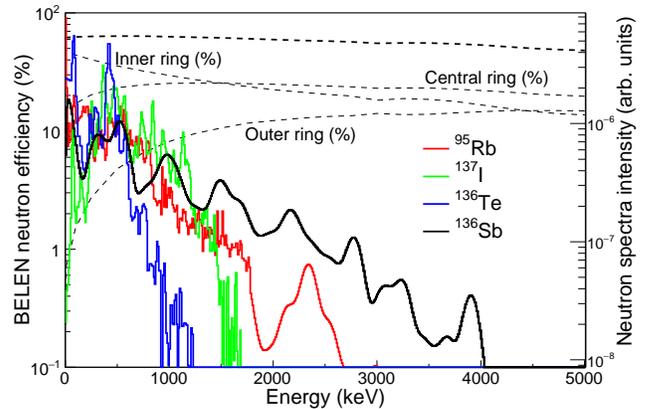}
\caption{BELEN detector efficiency from the GEANT4 simulation: total efficiency and for each one of the 3 rings of $^3$He proportional counters (dashed lines)~\cite{ariel2017private}, and neutron spectra of the measured isotopes (from ENSDF~\cite{nndcbnl}) as a function of the neutron energy (colored lines). See text and Table~\ref{table:BELEN-Charac} for details.}
\label{Fig:BELENEffnSpec}
\end{figure}

The version of the BELEN detector used in this experiment was specifically designed and optimized by means of Monte Carlo \textsc{MCNPX}~\cite{pelowitz2005mcnpx,fishman1996algorithms,rubinstein1981simulation,PhDRiego} and  \textsc{Geant4}~\cite{agostinelli2003geant4,ariel2017private} simulations in order to achieve a high and flat efficiency detection in the range from 0.1$-$2~MeV (see Fig.~\ref{Fig:BELENEffnSpec}). 
The most important constraint was in terms of efficiency: for the detection of two correlated neutrons, the efficiency, $\varepsilon_{2n}$, is roughly proportional to the square of the detection efficiency for a single neutron, $\varepsilon_{1n}$:
\begin{equation}
\centering
\bar{\varepsilon}_{2n} \approx \bar{\varepsilon}_{1n}^{2}
\label{Eq_Eff2}
\end{equation}
To avoid a low detection probability the central hole of BELEN was kept at 6~cm diameter, restricted to the diameter size of the vacuum tube of the implantation system. With this geometry, BELEN reached a one-neutron efficiency of $\approx$60\%, and $\approx$36\% for a two-neutron event, in the Q$_{\beta n}$ energy range of the isotopes of interest. The disadvantatge of this optimization compared to other designs was the impossibility of placing a high-purity germanium (HPGe) detector right behind the implantation point to verify the identity of the isotope implanted by $\gamma$-ray detection. This latter technique was used in the prior experiment during the same experimental campaign with another version of the BELEN detector, measuring isotopes in the same region of the nuclear chart~\cite{agramunt2017new}. The one-neutron efficiency for that version of BELEN detector was 40\% in the same energy range (0.1$-$2~MeV) due to a different geometry, while using the same number of $^{3}$He neutron counters. For the present work, the efficiency obtained in the simulations, shown in Fig.~\ref{Fig:BELENEffnSpec}, was experimentally validated at E$_{n}$=2.13~MeV using a $^{252}$Cf source, and with beams of isotopes with well-known P$_{1n}$ values, such as $^{95}$Rb and $^{137}$I, as reported in Section~\ref{sec:analysis}.\\

Previous experiments using the BELEN detector demonstrated that it works well with the self-triggered data acquisition system specifically implemented for BELEN, named \textit{GASIFIC}~\cite{agramunt2016characterization,caballero-folch2017beta}. This system integrates all signals from the $\beta$- and the $^{3}$He neutron counters recording their energy, and a time-stamp with a clock of 10~ns resolution to be able to build the $\beta$-neutron time-correlations over a certain time-window. In this experimental campaign, differential to single-ended converter modules, designed at JYFL, were added in the electronic chain. This made possible to link the output signals from the MPR-16-HV Mesytec preamplifiers, directly connected to the $^{3}$He tubes, to the SIS3316~\cite{StruckInc} sampling ADC modules in the data acquisition system~\cite{agramunt2017new}. This improvement enabled the recording of the data without requiring the signal shapers used in previous experiments.\\

\begin{figure}[!htbp]
\centering
\includegraphics[width = 1.0\columnwidth]{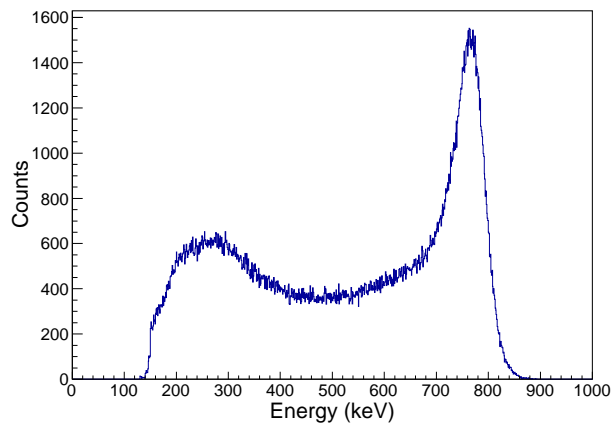}
\caption{BELEN energy spectrum acquired with the 48 $^3$He tubes for the $^{136}$Sb measurement.}
\label{fig:response}
\end{figure}

Regarding the response of the neutron detection system, the energy spectrum obtained for the neutron events comprises the range from a low-energy threshold at 191~keV up to the 764~keV peak. This energy is attributed to the kinetic energy released in the reaction:
\begin{equation}
\centering
^{3}\mathrm{He} + \mathrm{n} \longrightarrow~^{3}\mathrm{H} +~^{1}\mathrm{H} + 764~keV.
\label{E3_nReac}
\end{equation}
The 764~keV peak corresponds to the sum of the collection of the total energy released by the two reaction products, a triton and a proton. The lower energy detection threshold is related to the partial collection of the energy and the wall effect~\cite{ravazzani2006characterisation}. Gain-matching with a calibrated $^{252}$Cf source was carried out for all 48 $^{3}$He counters before the experiment. The stability of the overall detector response was checked regularly during the experiment. The accumulated spectrum for all 48 tubes during the $^{136}$Sb measurement is shown Fig.~\ref{fig:response}. The latter includes the uncorrelated neutron events, which were removed in the data analysis process together with other light particles and the noise at lower energies.
\section{Determination of the neutron branching ratios}\label{sec:analysis}
\begin{figure}[!ht]
\centering
\includegraphics[width = 1.0\columnwidth]{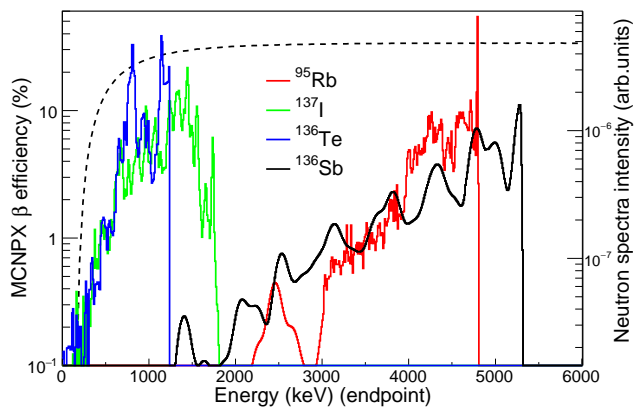}
\caption{The MCNPX simulation of the $\beta$ counter efficiency (dashed line), and electron endpoint energies from the neutron energy spectra of the measured isotopes (colored lines)~\cite{nndcbnl}. The lower $\beta$ efficiency at low energies affects the detection of correlated $\beta n$ events for those isotopes with a low Q$_\beta$ value (see text for details).}
\label{Fig:VICTOR}
\end{figure}

The equation that describes the total $\beta$-delayed neutron branching probability, P$_{n}$, of a nucleus is defined as the sum of all, single and multiple, neutron emission contributions present in the decay:
\begin{equation}
P_{n} = \sum_{x=1}^{N} P_{xn}
\label{Eq:PnSum}
\end{equation}
For nuclei with only one-neutron emission energetically allowed, i.e. Q$_{\beta n}>$0 and Q$_{\beta 2n}<$0, the P$_{n}$ value is equal to P$_{1n}$. This is the case for the $^{95}$Rb, $^{137}$I, and $^{136}$Te isotopes measured in this experiment. In order to determine their P$_{1n}$ values, we followed the methodology successfully implemented in a previous experiment with a similar setup at the IGISOL facility~\cite{agramunt2016characterization} in which the P$_{1n}$ values were deduced as:
\begin{equation}
P_{1n}=\frac{\bar{\varepsilon_{\beta}}}{\bar{\varepsilon}_{n}\bar{\varepsilon_{\beta}}^{'}}\frac{N_{\beta 1n}}{N_{\beta}}
\label{Eq:RatioCorr}
\end{equation}
where N$_{\beta 1n}$ is the number of the net $\beta$ and neutron time-correlated events, N$_{\beta}$ the number of $\beta$ decays registered, $\bar{\varepsilon_{\beta}}$ is the mean $\beta$ efficiency, $\bar{\varepsilon_{\beta}}^{'}$ the averaged $\beta$ efficiency above the S$_{n}$ weighted according to the Q$_{bn}$ and the neutron energy spectrum range (see Fig.~\ref{Fig:VICTOR}), and $\bar{\varepsilon}_{n}$ the neutron efficiency (see Fig.~\ref{Fig:BELENEffnSpec}).\\

\begin{figure*}[!ht]
\centering
\subfigure[$^{95}$Rb analysis]{\includegraphics[width = 1.0\columnwidth]{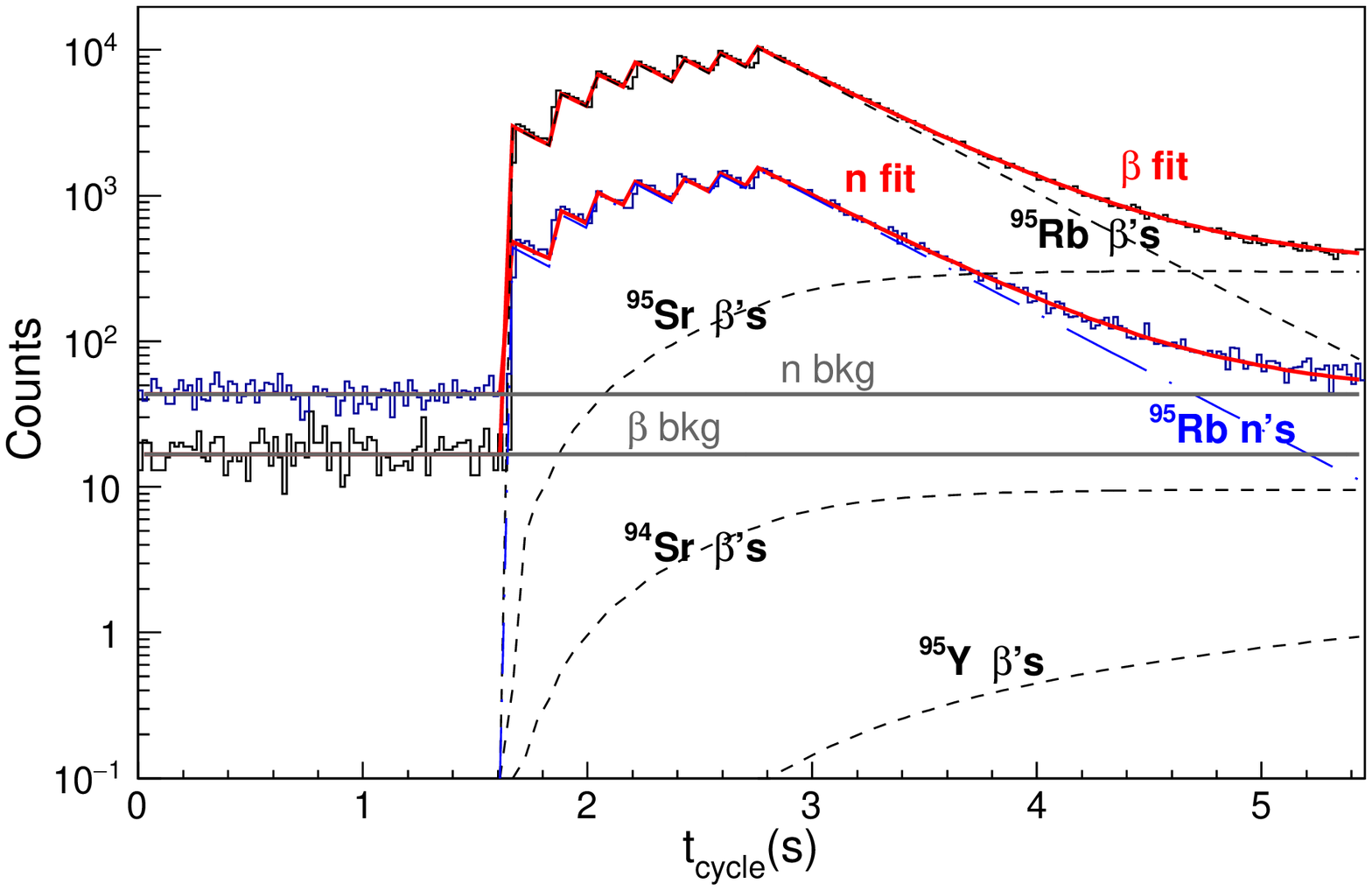}}
\subfigure[$^{137}$I analysis]{\includegraphics[width = 1.0\columnwidth]{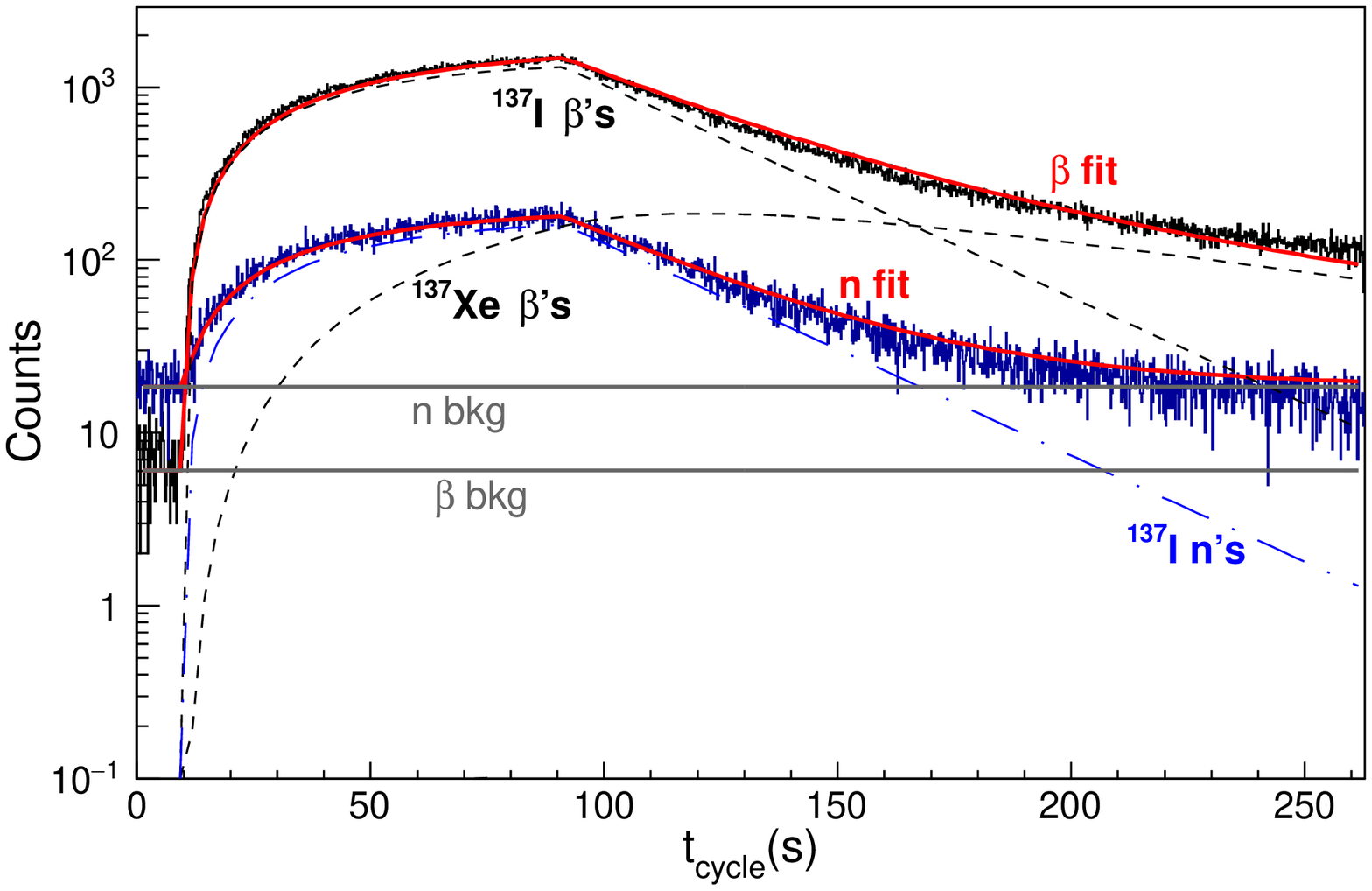}}
\subfigure[$^{136}$Te analysis]{\includegraphics[width = 1.0\columnwidth]{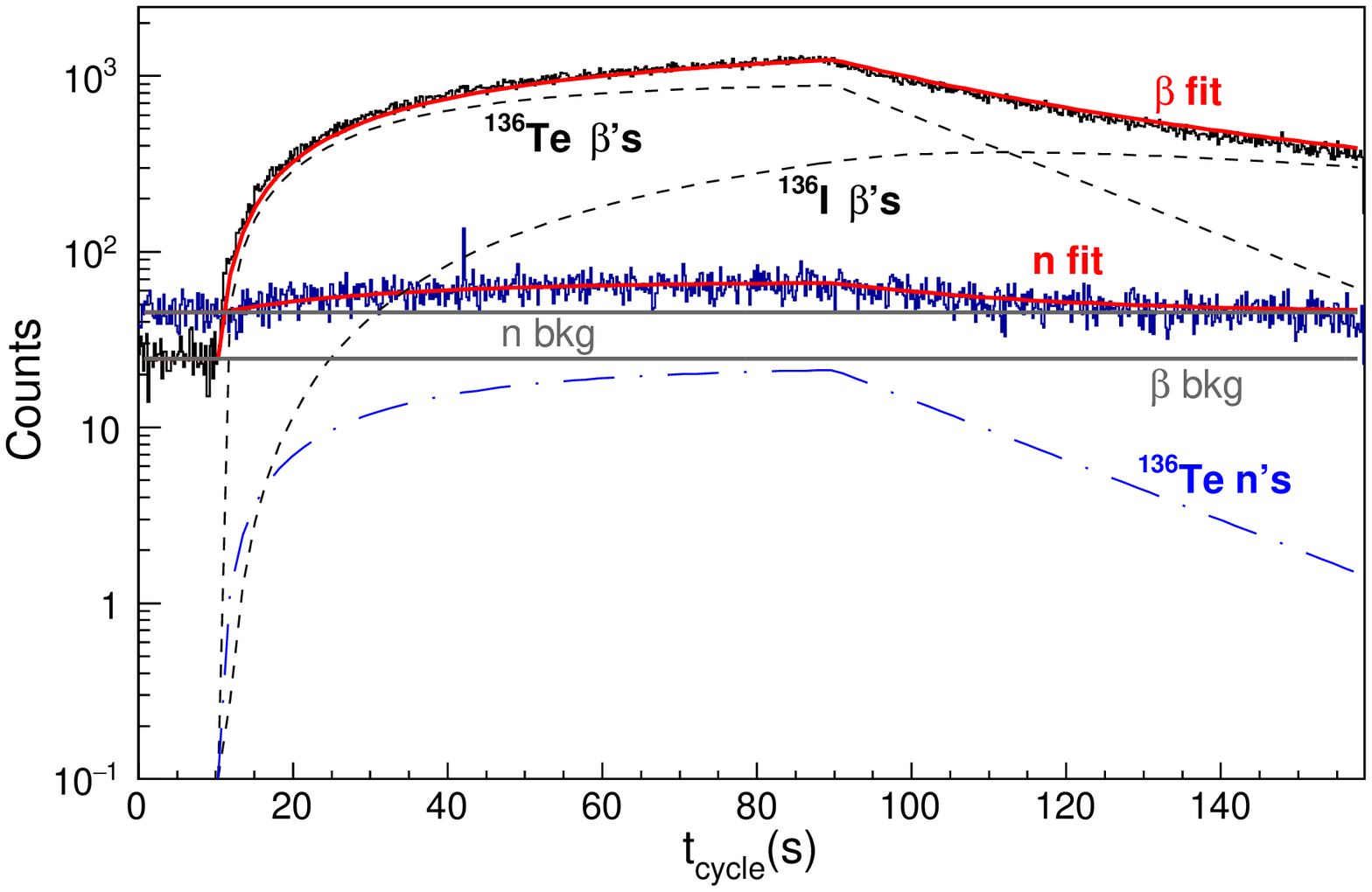}}
\subfigure[$^{136}$Sb analysis]{\includegraphics[width = 1.0\columnwidth]{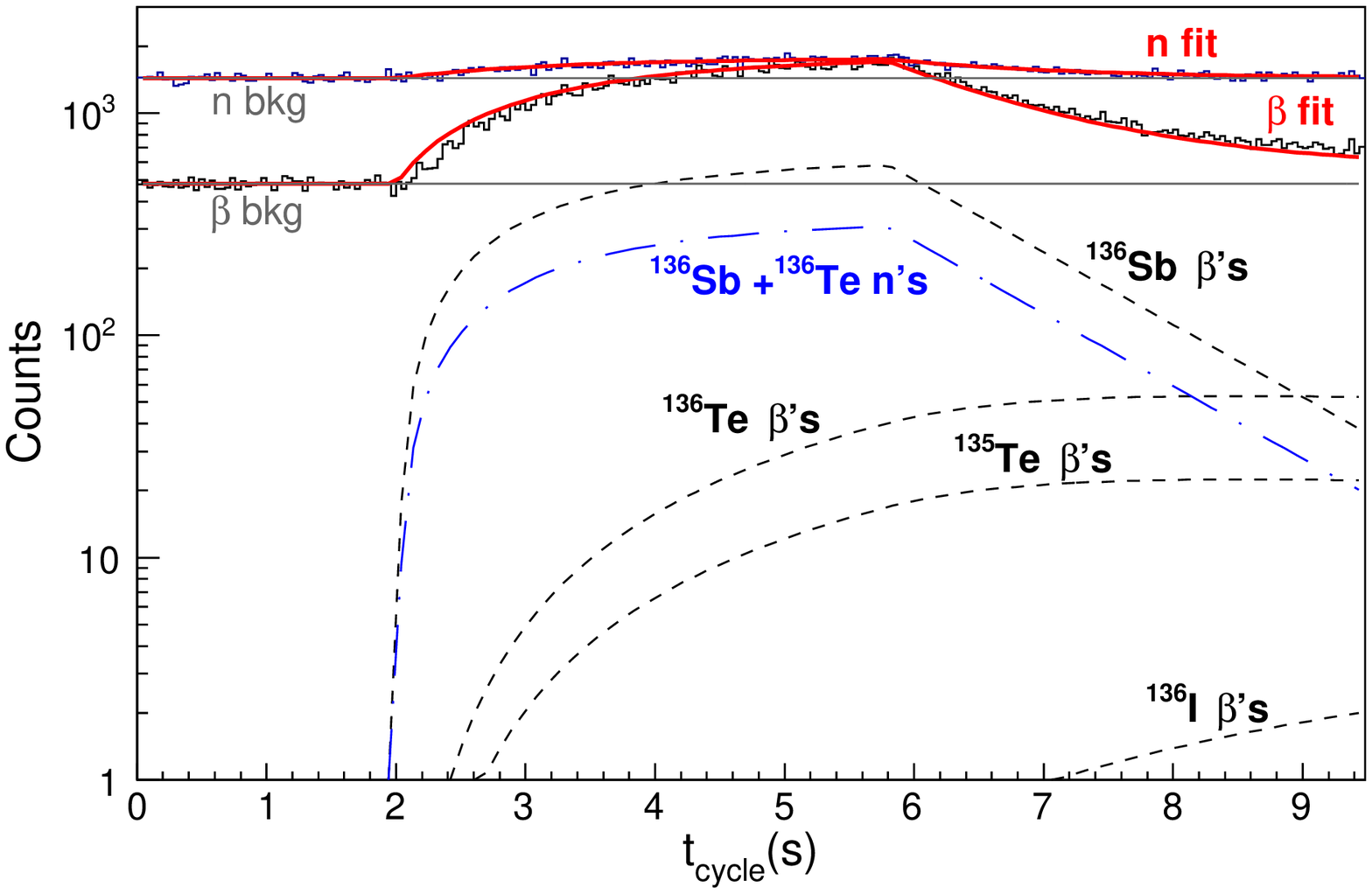}}
\caption{Analysis of the $\beta$ and neutron events accumulated during the implantation-and-decay cycles in the growth-and-decay curves for each isotope measured. In the analysis of $^{137}$I (b) the fit function also includes a parameter to reflect the loss of the xenon isotopes in the chain. As xenon is a noble gas, a fraction of the nuclei rapidly escape from the implantation tape. Details of this effect are presented in Ref.~\cite{agramunt2016characterization}. In the case of $^{95}$Rb (a) the trap purification cycle can be identified. In all cases the ions were extracted from the trap in bunches, once in every 181~ms according to the length of this cycle. This was repeated as many times as needed during the ``beam on" time. So the beam was not continuous but rather a quasi-continuous.}
\label{Fig:Fits}
\end{figure*}

The analysis of the acquired data directly provides three parameters that are needed to evaluate and determine the neutron branching ratios. These are the number of $\beta$ particles and the number of neutrons detected, together with the $\beta n$ time-correlation events. Figure~\ref{Fig:Fits} shows the analysis to determine the $\beta$ and neutron integrals for each measured isotope, $^{95}$Rb, $^{137}$I, $^{136}$Te, and $^{136}$Sb, from the growth-and-decay curves, using the Bateman equations~\cite{Bateman1910equations}.
 The particular growth (implantation) and decay times in the curves were precisely set according to the half-lives, taken form the literature~\cite{nndcbnl}, of the isotopes present in the decay chain. The flat area at the beginning of the timing cycle (see fig.~\ref{Fig:Fits}) corresponds to the characterization of the $\beta$ and neutron background. Table~\ref{Table:GDcurves} provides the Q$_{\beta}$ values of interest together with half-lives and the details of their implantation and decay cycle times. The numerical results from Fig.~\ref{Fig:Fits} for the precursor isotopes are shown in later tables.\\
\begin{table}[ht!]
\caption{Half-lives~\cite{nndcbnl} and Q$_{\beta}$ values~\cite{audi2017nubase2016} of the investigated isotopes and characteristics of the implantation-and-decay cycle, corresponding to the growth-and-decay curves (see Fig.~\ref{Fig:Fits}).\label{Table:GDcurves}}
\begin{ruledtabular}
\begin{tabular}{lcccc}
\\
Nuclei& $^{95}$Rb& $^{137}$I& $^{136}$Te& $^{136}$Sb\\
\hline
\\
t$_{1/2}$(s)         &0.3777(8)& 24.5(2)& 17.63(8)& 0.923(14)\\
Q$_{\beta^-}$(keV)   & 9228(20)& 6027(8)& 5120(14)& 9916(7)\\ 
Q$_{\beta n}$(keV)   & 4881(20)& 2002(8)& 1292(6) & 5147(7)\\
Q$_{\beta 2n}$(keV)  &$-$1951(20)&$-$6082(8)&$-$6509(15)& 1884(6)\\
Impl. time (s)	     &     1.27&    81.3&     79.5&   3.80\\
Decay time (s)       &     2.54&   171.5&     68.5&   3.71\\
\end{tabular}
\end{ruledtabular}
\end{table}
In the present work $^{137}$I and $^{95}$Rb beams were used as calibration isotopes because of their well-known P$_{1n}$ values~\cite{rudstam1993delayed,pfeiffer2002status,iaea2018evaluation}. In addition, these two isotopes have very different neutron energy spectra (see Fig.~\ref{Fig:BELENEffnSpec}). In the case of $^{95}$Rb, its neutron energy spectrum is mainly centered at lower energies, with a large Q$_{\beta n}$ value of 4.8~MeV. This indicates that the lower $\beta$ detection probability at lower energies (see Fig.~\ref{Fig:VICTOR}) does not miss $\beta n$ correlated events in the analysis. Therefore, assuming $\bar{\varepsilon_{\beta}}\approx\bar{\varepsilon_{\beta}}^{'}$ in Eq.~\ref{Eq:RatioCorr}, it is possible to determine its P$_{1n}$ value to a first approximation with the equation:
\begin{equation}
P_{1n}\simeq\frac{1}{\bar{\varepsilon}_{n}}\frac{N_{\beta 1n}}{N_{\beta}}
\label{E-Pn-1}
\end{equation}
The resulting P$_{1n}$ value for $^{95}$Rb, considering a constant $\bar{\varepsilon_{n}}$ value of 62.98\%, taken from the simulation and weighted with the neutron spectrum, is 8.6(4)\%, in agreement with the literature value of 8.94(37)\%~\cite{iaea2018evaluation}. On the other hand, $^{137}$I has a neutron spectrum spread over the whole energy range up to its endpoint at $\approx$2~MeV (see Fig.~\ref{Fig:BELENEffnSpec}). In this case its analysis is affected due to the non-constant $\beta$ efficiency at low energies (see Fig~\ref{Fig:VICTOR}), and the simplified expression detailed in Eq.~\ref{E-Pn-1} is not accurate enough. Thus, we defined an expression to avoid the $\bar{\varepsilon_{\beta}}$ and $\bar{\varepsilon_{\beta}}^{'}$ dependence in Eq.~\ref{Eq:RatioCorr} as in Ref.~\cite{agramunt2016characterization}, using the total number of detected neutrons, N$_{n}$, independently of the coincidence or not with the $\beta s$:
\begin{equation}
P_{1n}=\frac{\bar{\varepsilon}_{\beta}}{\bar{\varepsilon}_{n}}\frac{N_{n}}{N_{\beta}},
\label{Eq:Ratio}
\end{equation}
from which, it is then possible to determine the $\bar{\varepsilon_{n}}/\bar{\varepsilon_{\beta}}$ ratio with the well-known $P_{1n}$ values of the $^{95}$Rb and $^{137}$I calibration isotopes. The ratio obtained can be used to determine the remaining P$_{1n}$ values in this experiment. Table~\ref{Table:EnEbratio} summarizes the integral values of the $\beta n$ correlated events, the number of $\beta$ events, N$_{\beta}$, and the number of neutrons, N$_{n}$, for the $^{95}$Rb and $^{137}$I measurements, together with the calculated $\bar{\varepsilon}_{n}$/$\bar{\varepsilon}_{\beta}$ ratio.\\

\begin{table}[ht!]
\caption{The experimental values measured for $^{95}$Rb and $^{137}$I and their respective $\bar{\varepsilon}_{\beta}$/$\bar{\varepsilon}_{n}$ ratio.\label{Table:EnEbratio}}
\begin{ruledtabular}
\begin{tabular}{cccccc}
Nuclei&P$_{1n}$(\%)& N$_{\beta n}$&N$_{\beta}$&N$_{n}$&$\bar{\varepsilon}_{n}$/$\bar{\varepsilon}_{\beta}$\\
\hline
\\
$^{95}$Rb& 8.94(37)~\cite{iaea2018evaluation}& 33011& 610229& 90445&1.66(7)\\
$^{137}$I& 7.66(14)~\cite{iaea2018evaluation}& 21888\footnote{Biased value due to the lower $\bar{\varepsilon}_{\beta}$ at lower energies.}& 592009& 72031& 1.59(3)\\
\end{tabular}
\end{ruledtabular}
\end{table}
The average of the $\bar{\varepsilon}_{n}$/$\bar{\varepsilon}_{\beta}$ ratio, calculated from $^{95}$Rb and $^{137}$I measurements in Table~\ref{Table:EnEbratio}, is 1.62(7). As these two isotopes have large Q$_{\beta}$ values this ratio value is expected to be alike. Thus, we assume that this value also applies for all the other isotopes measured in this experiment. 

\subsection{The P$_{1n}$ value of $^{136}$Te}\label{subsec:Te}
\begin{figure}[!ht]
\centering
\includegraphics[width = 1.0\columnwidth]{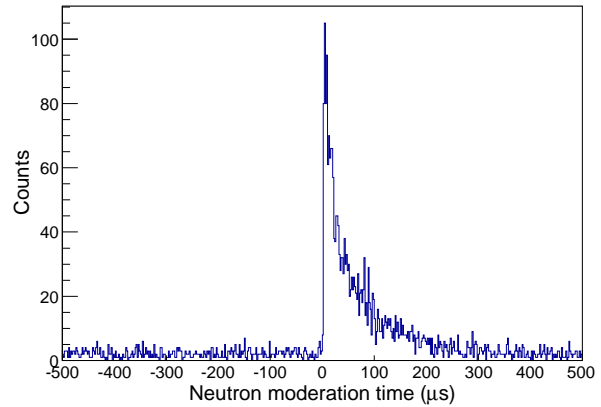}
\caption{$\beta 1n$ correlation events for the $^{136}$Te analysis. The 2630 forward and 548 backward events were registered during the experiment. According to the $\beta n$ background level on the left side, the neutron moderation time in the polyethylene for this experiment was determined to be 500~$\mu$s.}
\label{Fig:136TeBN}
\end{figure}

With a Q$_{\beta n}$ value of 1292(6)~keV, $^{136}$Te has an energy window that allows $\beta$-delayed one-neutron emission. In this case the expected energy spectrum of the emitted neutrons is also affected by the non-constant $\bar{\varepsilon}_{\beta}$ at low energies described above for $^{137}$I. Thus, the equation to determine its neutron branching ratio can be defined by Eq.~\ref{Eq:Ratio}, using the $\bar{\varepsilon}_{n}$/$\bar{\varepsilon}_{\beta}$ ratio determined with the calibration isotopes. With the integral values obtained in the analysis of the growth-and-decay curves in Fig.~\ref{Fig:Fits}, the analysis yields a P$_{1n}$ of 1.47(6)\%. This value is slightly higher but in fair agreement with those reported in the literature and the IAEA evaluation: 1.31(5)\%~\cite{rudstam1993delayed,iaea2018evaluation} and 1.26(20)\%~\cite{pfeiffer2002status}. Using the simplified Eq.~\ref{E-Pn-1}, with the $\beta n$ correlated detected events derived from the analysis of Fig.~\ref{Fig:136TeBN},
 the P$_{1n}$ value would be 0.90(5)\%. Comparing this value with the one obtained using the efficiency ratio correction we observe an increase of about 63\%. This gives an idea of the fraction of correlated events missed because of the lower $\bar{\varepsilon}_{\beta}$ at low energies when measuring an isotope with a neutron energy spectrum such as $^{136}$Te (see Fig.~\ref{Fig:VICTOR}). The acquired data for this isotope in this measurement were recorded for 165 cycles which included a background measurement of 10~s, an implantation time of 79.5~s, and a decay time of 68.5~s for each one, giving 158.5~s per cycle (see Fig.~\ref{Fig:Fits} and Table~\ref{Table:GDcurves}), i.e. 7.3~h of beamtime. A total of 2082 net $\beta 1n$ correlated events were registered for this isotope. The values of this analysis are reported in Table~\ref{Table:Results1}.
\begin{table}[ht!]
\caption{$^{136}$Te and $^{136}$Sb neutron emission probabilitiy results.\label{Table:Results1}}
\begin{ruledtabular}
\begin{tabular}{ccccccc}
Nuclei& N$_{\beta 1n}$& N$_{\beta 2n}$ & N$_{\beta}$&  N$_{n}$& P$_{1n}$(\%)& P$_{2n}$(\%)\\
\hline
$^{136}$Te& 2082      &  - & 360655 & 8645        & 1.47(6)& -\\
$^{136}$Sb& 9328& 21.8 & 57590& 30455\footnote{Includes the 1n and 2n events from $^{136}$Sb, and 1n events from $^{136}$Te.}& 32.2(15)& 0.14(3)\\
\end{tabular}
\end{ruledtabular}
\end{table}

\subsection{P$_{1n}$ and P$_{2n}$ values of $^{136}$Sb}
As described in Section~\ref{sec:intro}, when the state populated after the $\beta$ decay is higher than the two-neutron separation energy, S$_{2n}$, in the daughter nucleus (Q$_{\beta 2n} > 0$), double neutron emissions are allowed. This is the case for $^{136}$Sb with Q$_{\beta 2n}$=1884(6)~keV. Its measurement in this experiment comprised 26465 recorded implantation-and-decay cycles of 9.51~seconds ($\approx$3~days). As can be observed in Fig.~\ref{Fig:Fits}, in each cycle the first 2~s were used to characterize the $\beta$ and neutron background. Table~\ref{Table:GDcurves} details the implantation-and-decay cycle times. The neutron branching ratios of this isotope have two contributions to the P$_{n}$ (Eq.~\ref{Eq:PnSum}): the P$_{1n}$ and the P$_{2n}$ values.

\begin{figure}[!ht]
\centering
\includegraphics[width = 1.0\columnwidth]{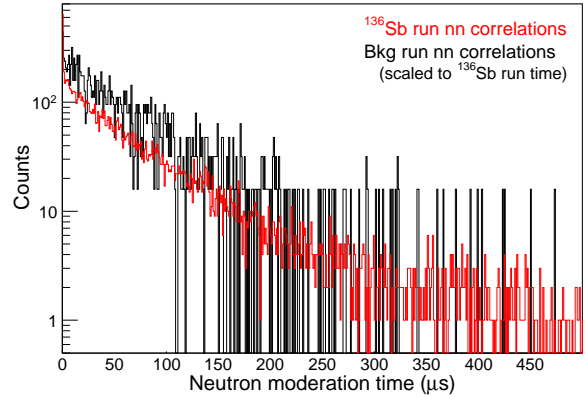}
\caption{(Color online) Background level evaluation of double neutron events, N$_{2n}$, for the $^{136}$Sb setting. As can be observed, the number of events in a background measurement is of the same order as in the setting of interest. See text for details.}
\label{Fig:136SbNNCorr}
\end{figure}

In order to determine its neutron branching ratios, several steps are necessary. Eq.~\ref{Eq:Nn} describes the total number of neutrons, N$_{n}$, after background subtraction:
\begin{equation}
N_{n} = N_{1n}(^{136}\mathrm{Sb}) + 2 N_{2n}(^{136}\mathrm{Sb}) + N_{1n}(^{136}\mathrm{Te}).
\label{Eq:Nn}
\end{equation}
As can be seen the total number of neutrons has contributions from the one- and two-neutron emission branching ratios of $^{136}$Sb plus a small contribution from its daughter $^{136}$Te. The latter is present in the decay chain and the relevant number of neutrons can be determined with the information extracted from the $^{136}$Sb $\beta$-decay analysis (Fig.~\ref{Fig:Fits}) and its P$_{1n}$ value. In order to calculate the contributions of one-neutron and two-neutron events from the $^{136}$Sb decay in Eq.~\ref{Eq:Nn}, we define the following expressions:
\begin{equation}
N_{1n(^{136}Sb)} = \bar{\varepsilon}_{n} P_{1n} \frac{N_{\beta}}{\bar{\varepsilon}_{\beta}} + 2 \times\bar{\varepsilon}_{n} (1-\bar{\varepsilon}_{n}) P_{2n} \frac{N_{\beta}}{\bar{\varepsilon}_{\beta}}
\label{Eq:N1n}
\end{equation}
and
\begin{equation}
N_{2n(^{136}Sb)} = \bar{\varepsilon}_{n}^{2} P_{2n} \frac{N_{\beta}}{\bar{\varepsilon}_{\beta}}
\label{Eq:N2n}
\end{equation}
These expressions relate the number of $\beta$ events, N$_{\beta}$, and neutron events, N$_{xn}$, and their detection efficiency, to the unknown P$_{1n}$ and P$_{2n}$ values. However, from the available information, the analysis only provides the net number of N$_{\beta}$ events for each one of the species in the decay chain, and the net number of the total neutron events, N$_{n}$, after background subtraction. Firstly, we attempted to determine the net number of two-neutron correlations, N$_{2n}$, from the decay of $^{136}$Sb subtracting the two-neutron correlations determined in a background run. As shown in Fig.~\ref{Fig:136SbNNCorr}, this method is not useful because of the large uncertainty in the background subtraction.
The large background can be associated with different kinds of interactions between particles, either cosmic rays or reactions of the beam with materials, and the BELEN detector.\\

\begin{figure}[!ht]
\centering
\includegraphics[width = 1.0\columnwidth]{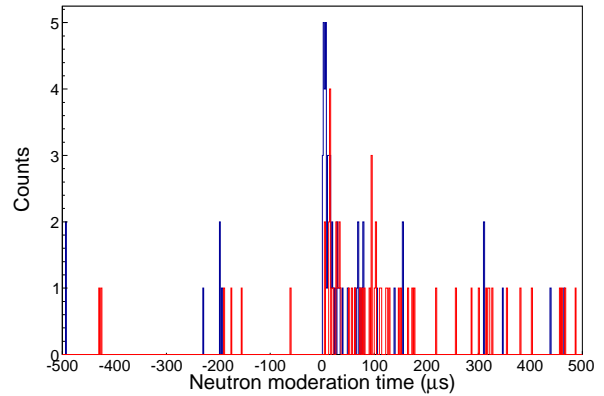}
\caption{(Color online) $^{136}$Sb $\beta 2n$ correlation events. In blue are shown the first correlated neutrons after a $\beta$ decay, in red the second neutrons in the correlation. A total of 55 forward events were recorded, in which are included several background contributions (see text for details).}
\label{Fig:136Sbb2n}
\end{figure}

\begin{figure}[!ht]
\centering
\includegraphics[width = 1.0\columnwidth]{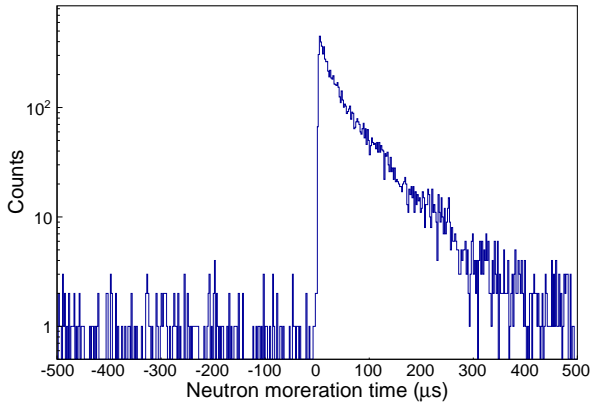}
\caption{$\beta 1n$ correlation for the $^{136}$Sb analysis. A total of 9494 forward and 168 backward events were registered. Beyond the calculation to determine the number of $\beta 1n$ net events, the background rate in this $\beta 1n$ channel is relevant in the determination of the truly $\beta 2n$ events. (See text for details).}
\label{Fig:136Sbb1n}
\end{figure}

Therefore, the strategy to determine the neutron branching ratios was based on the evaluation of the direct $\beta 2n$ and $\beta 1n$ correlation events recorded in the data analysis. Figures~\ref{Fig:136Sbb2n} and~\ref{Fig:136Sbb1n} show, respectively, the $\beta 2n$ and $\beta 1n$ time-correlation events registered within a neutron moderation time-window of 500~$\mu s$.
In the analysis of the $\beta 2n$ correlation were registered 55 forward events, $N_{\beta 2n}^{f}$, and 6 backward events, $N_{2n\beta}$, corresponding to the number of two accidental neutrons in the neutron moderation time-window. In order to evaluate the total background contributions present in the 55 $\beta 2n$ detected events, we assume that the neutron rates are the same within the neutron moderation time-window, before and after a $\beta$ decay. For the case of two neutrons accidentally correlated per detected $\beta$, we define the parameter $r_{2n}$ as the number of backward time-correlations, $N_{2n\beta}$, and the total number of $\beta$-decays, $N_{\beta}$, obtained from Fig.~\ref{Fig:Fits} and reported in Table~\ref{Table:Results1}:
\begin{equation}
r_{2n} = \frac{N_{2n\beta}}{N_{\beta}}.
\label{Eq:r2}
\end{equation}
The amount of events contributing due to this $r_{2n}$ rate includes all $\beta$ decays except those coming from the $\beta 1n$ and $\beta 2n$ events, i.e. $r_{2n}\times(N_{\beta}-N_{\beta 1n}-N_{\beta 2n})$. However, according to the measured values, we can assume that $N_{\beta} >> N_{\beta 1n} >> N_{\beta 2n}$, and the background from this $\beta$-$2n$ channel is approximately of $r_{2n} N_{\beta}$. Another relevant background contribution comes from the detection of one accidental neutron inside the same correlation time-window of a truly correlated neutron. In this case we define the parameter $r_{1n}$ as the $\beta 1n$ backward time-correlations, $N_{1n\beta}$, as obtained in the analysis presented in Fig.~\ref{Fig:136Sbb1n}, and the total number of $\beta$-decays $N_{\beta}$ from Fig.~\ref{Fig:Fits}:
\begin{equation}
r_{1n} = \frac{N_{1n\beta}}{N_{\beta}}.
\label{Eq:r1}
\end{equation}
In addition, this $\beta$-$1n$ background channel has also contributions from those neutrons in the $\beta$-$2n$ channel in which one of the neutrons has not been detected. This can be estimated with the neutron efficiency factor defined as:
\begin{equation}
r_{\epsilon} = \frac{(1-\epsilon_{n})}{\epsilon_{n}},
\label{Eq:re}
\end{equation}
being the total contribution of the one neutron channel as $r_{1n}\times(N_{\beta 1n}+2\times r_{\varepsilon} N_{\beta 2n})$.
To summarize, Eqs.~\ref{Eq:b1nm} and~\ref{Eq:b2nm} describe the system that relates the number of measured $\beta 1n$ and $\beta 2n$ (forward) events, $N_{\beta 1n}^{f}$ and $N_{\beta 2n}^{f}$, with the true events, $N_{\beta 1n}$ and $N_{\beta 2n}$, and background contributions, assuming $N_{\beta} >> N_{\beta 1n} >> N_{\beta 2n}$:
\begin{equation}
N_{\beta 1n}^{f} \approx (1-r) (N_{\beta 1n}+2r_{\epsilon}N_{\beta 2n}) + r_{1n} N_{\beta}
\label{Eq:b1nm}
\end{equation}
\begin{equation}
N_{\beta 2n}^{f} \approx (1-r) N_{\beta 2n} + r_{2n}N_{\beta} + r_{1n}(N_{\beta 1n}+2r_{\epsilon}N_{\beta 2n})
\label{Eq:b2nm}
\end{equation}
where $r$ is the sum of all accidental neutrons per detected $\beta$, $r=r_{1n}+r_{2n}+r_{3n}...$, which are added on each particular background channel in the latter equations. In this measurement, this sum was simplified considering that $r_{1n}>>r_{2n}>>r_{3n}$, and only the first two contributions were taken into account. Solving the system presented in Eqs.~\ref{Eq:b1nm} and~\ref{Eq:b2nm}, the resulting values for the $N_{\beta 2n}$ and $N_{\beta 1n}$, in the $^{136}$Sb measurement, are 21.8 and 9328 respectively. The method described above is presented in detail in Ref.~\cite{tolosa2018commissioning} within the BRIKEN Project~\cite{dillmann2018beta}.\\

If we were to assume that the mean energy of the neutron spectrum is not affected by the lower $\varepsilon_{\beta}$ at low energies, the P$_{2n}$ value could be calculated with a simplified expression such as Eq.~\ref{Eq:P2ns}, analogously to Eq.~\ref{E-Pn-1} used for the case of $^{95}$Rb for its P$_{1n}$ value:
\begin{equation}
P_{2n} = \frac{1}{\bar{\varepsilon}_{n}^{2}} \frac{N_{\beta 2n}}{N_{\beta}}
\label{Eq:P2ns}
\end{equation}
With this assumption, the P$_{2n}$ value would be 0.10(2)\%. However, due to the expected low $\bar{\varepsilon}_{\beta}$ in the energy region of interest for the two-neutron energy spectrum (Q$_{\beta 2n}$ value $=$ 1884(6)~keV, see Fig.~\ref{Fig:VICTOR}), it is necessary to apply the correction based on the $\bar{\varepsilon_{\beta}}^{'}$ (Eq.~\ref{Eq:P2nsCorr}) in the same way as Eq.~\ref{Eq:RatioCorr} for the case of one-neutron emission (see Section~\ref{sec:analysis}):
\begin{equation}
P_{2n} = \frac{1}{\bar{\varepsilon}_{n}^{2}} \frac{\bar{\varepsilon}_{\beta}}{\bar{\varepsilon_{\beta}}^{'}} \frac{N_{\beta 2n}}{N_{\beta}}
\label{Eq:P2nsCorr}
\end{equation}
The unknown neutron energy spectrum for the $^{136}$Sb two-neutron emission suggests the use of the calculated $\bar{{\varepsilon}_{\beta}}^{'}$ value for $^{137}$I, as its Q$_{\beta n}$ value, 2002~keV, is very close to the $^{136}$Sb Q$_{\beta 2n}$ value. So, using the $\bar{\varepsilon}_{n}$/$\bar{\varepsilon}_{\beta}$ coefficient, 1.62(7), previously determined, the P$_{2n}$ value is calculated to be 0.14(3)\%. Having obtained the P$_{2n}$, the P$_{1n}$ value, has been determined to be 32.2(15)\%.

\section{Summary and discussion}\label{sec:discussion}
The P$_{1n}$ and P$_{2n}$ values obtained in this work are summarized in Table~\ref{Table:Discussion} together with the previously available experimental values and theoretical predictions from several models, including the finite-range droplet model plus the quasiparticle random-phase approximation (FRDM+QRPA)~\cite{moller2003new}, the density functional theory plus continuum QRPA plus relativistic QRPA(DF3+cQRPA+RQRPA)~\cite{borzov2006beta,borzov2012ICFN5}, the Koura-Tachibana-Uno-Yamada (KTUY)~\cite{koura2005nuclidic}, the relativistic Hartree-Bogoliubov plus RQRPA (RHB+RQRPA)~\cite{marketin2016large}, the QRPA plus Hartree-Fock (QRPA-HF)~\cite{moller2018nuclear}, and the semi-empirical effective density model~\cite{miernik2014beta}. 
\begin{table*}[ht!]
\caption{Results\label{Table:Discussion}}
\begin{ruledtabular}
\begin{tabular}{ccccccc|cccccc}
&&&&&&&&&&&&\\
Isotope& t$_{1/2}$~(s)~\cite{nndcbnl}& P$_{1n}$(\%)&P$_{2n}(\%)$&Technique& Date and&& & t$_{1/2}$(s) & P$_{1n}$(\%)&P$_{2n}$(\%)&Model&Ref.\\
       &      \multicolumn{4}{c}{(Experimental)}		        &Ref.&& & \multicolumn{5}{c}{(Theory)}\\
\hline
&&&&&&&&&&&&\\
$^{136}_{52}$Te$^{84}$	& 17.63(8)&  1.47(6)& -&$\beta$,n&(This work) 	             & & & 10.166&  2.43& 0.0& FRDM+QRPA&~\cite{moller2003new}\\
			            &         & 1.26(20)& -&$\beta$,n&(2002)~\cite{pfeiffer2002status} & & & 21.189&  1.80&   -& DF3+cQRPA+RQRPA&~\cite{borzov2012ICFN5}\\
			            &         &  1.31(5)& -&$\beta$,n&(1993)~\cite{rudstam1993delayed} & & &  0.600&  0.41&   -& KTUY&~\cite{koura2005nuclidic}\\
			            &         &  1.7(8)\footnote{\label{footsecond}Updated values from the evaluation performed in Ref.~\cite{rudstam1993delayed}.}& -&Fission,$\beta$,n&(1978)~\cite{crancon1978half} & & &  0.548&   0.4& 0.1& RHB+RQRPA&~\cite{marketin2016large}\\
			            &         &   0.7(4)& -&Fission,n&(1977)~\cite{rudolph1977half}       & & & -& 2& 0& QRPA-HF&~\cite{moller2018nuclear}\\
			            &         &  1.31(5)& -&Evaluation&(2018)~\cite{iaea2018evaluation}& & &      -&   2.8& 0.0& Semi-empirical&~\cite{miernik2014beta}\\			            
\hline
$^{136}_{51}$Sb$^{85}$	& 0.923(14)&32.2(15)&0.14(3)&$\beta$,n&(This work)           & & & 1.998&     33.5&       6.19& FRDM+QRPA&~\cite{moller2003new}\\
			&   &        -&1.4(2)\footnote{\label{footfirst}Measurement with $A=136$ isobaric contamination.}&$\beta$,n&(2011)~\cite{TESTOV-EspRus2011} & & &   0.8& 10.5(51)& 4.15(1.05)& DF3+cQRPA+RQRPA&~\cite{borzov2016self}\\
			&	& 19.5(18)&     -&$\beta$ recoil&(2015)~\cite{Caldwell2015PhD}             & & & 0.760&    33.20&        0.0& KTUY&~\cite{koura2005nuclidic}\\
			&	& 23.2(68)&     -&$\beta$,n&(2002)~\cite{pfeiffer2002status}               & & & 0.175&      3.8&        0.2& RHB+RQRPA&~\cite{marketin2016large}\\
			&	& 16.3(32)&     -&$\beta$,n&(1993)~\cite{rudstam1993delayed}               & & &     -&     30.0&        0& QRPA-HF&~\cite{moller2018nuclear}\\
			&	& 33(40)\footref{footsecond}& -&Fission,$\beta$,n&(1978)~\cite{crancon1978half} & & &     -&     37.3&        0.0& Semi-empirical&~\cite{miernik2014beta}\\
			&	& 44(57)\footref{footsecond}& -&Fission,n&(1977)~\cite{rudolph1977half}       & & &  0.46&     17.1&       0.28& Microscopic Finite &~\cite{lyutostansky1983estimation}\\
			&	& 18.7(18)&    $<$1\footnote{\label{footfirst}Based on preliminary results of this study.}&Evaluation&(2018)~\cite{iaea2018evaluation}            & & &      &         &           & Fermi-system theory&\\
&&&&&&&&&&&\\
\end{tabular}
\end{ruledtabular}
\end{table*}
The P$_{1n}$ value obtained for $^{136}$Te, 1.47(6)\%, is higher but in fair agreement with those reported in Refs.~\cite{rudstam1993delayed,pfeiffer2002status} and~\cite{iaea2018evaluation}. Concerning $^{136}$Sb, the analysis yields a P$_{1n}$ value of 32.2(15)\%. This is higher than previous experimental results also taken from Refs.~\cite{rudstam1993delayed,pfeiffer2002status} and~\cite{iaea2018evaluation}. In the case of the P$_{2n}$ we have obtained a value of 0.14(3)\%. This is one order of magnitude lower than the 1.4\% estimated in a measurement with isobaric contamination~\cite{TESTOV-EspRus2011}, and the predictions of the FRDM+QRPA and the DF3+cQRPA+RQRPA models.\\

As shown in Table~\ref{Table:Discussion}, the values of most of the theoretical predictions are far from the experimental values. Some of them are compatible for $^{136}$Sb but not for $^{136}$Te, and vice versa. With the current results it is not easy to decide which of the models reproduces better the experimental results. The DF3+cQRPA~\cite{borzov2012ICFN5,borzov2016self} approaches better the magnitudes of the P$_{1n}$ and half-life for $^{136}$Te, but in the case of $^{136}$Sb, although the predicted half-life is close to the experimental value, it underestimates P$_{1n}$ value and overestimates the P$_{2n}$ value. The model which better reproduces the $^{136}$Sb data is the KTUY~\cite{koura2005nuclidic} but the half-life predicted for $^{136}$Te remains far from the experimental value. Regarding the QRPA-HF~\cite{moller2018nuclear} model, it is the one which better reproduces the P$_{1n}$ and P$_{2n}$ values for both isotopes according the obtained experimental results.\\

To date, several calculations and estimates for multiple-neutron emission beyond A=100 have also been reported. Table~\ref{Table:Discussion} reports the available ones for the isotopes measured in this study. Some of them follow the main theoretical models describing the strength functions and some others are estimates and extrapolations like~\cite{miernik2014beta}. Concerning experimental measurements, there is a recent P$_{2n}$ value reported for $^{140}$Sb~\cite{moon2017nuclear}. This value was determined through an indirect measurement based on relative intensities of $\gamma$-rays observed from transitions that were identified as belonging to $^{138}$Te. The value reported is \textit{``about" 8\%} and no uncertainty was given. This suggests again the need for direct neutron measurements in order to obtain more precise data on neutron emission branching ratios.

\section{Conclusions and outlook}\label{sec:conclusions}
We have determined the neutron branching ratios for $^{136}$Te and $^{136}$Sb through a direct neutron measurement. This represents the first experimental multiple-neutron emission value above A=100 and an improvement for the values available so far for these isotopes. As discussed above, some of the theoretical predictions agree well with the values obtained, but none shows agreement for all of the parameters present in the decay. This, together with the discrepancies with the experimental data available, indicates that more measurements with pure beams and direct neutron detection are needed in order to provide further input for the models in this region, and to study the nuclear properties above the neutron separation energies. Consequently, this experimental campaign in which six more isotopes included in the IAEA priority list for reactor physics~\cite{IAEA2014INDC} were also measured~\cite{agramunt2017new}, and present and future campaigns like BRIKEN~\cite{dillmann2018beta} are an opportunity to increase the amount of data available in this field.\\
 
It is also of interest to determine whether the two neutrons are emitted simultaneously or sequentially in the $\beta$-delayed two neutron-emission process. In the first case an angular correlation between both neutrons and the $\gamma$-ray emitted by the final nucleus would be expected. In the second case it would be necessary to correlate these events with very narrow time-windows to be able to confirm the phenomenon. In this work we were not able to address this question because of the moderation of the neutrons in the polyethylene matrix and the statistics available.

\begin{acknowledgments}
This work was supported by the National Research Council of Canada (NSERC) Discovery Grants SAPIN-2014-00028 and RGPAS 462257-2014 at TRIUMF, and by the \textit{Spanish Ministerio de Economia y Competitividad} under grants: FPA2010-17142, AIC-D-2011-0705, FPA2011-28770-C03-03, FPA2011-24553, FPA2014-52823-C2-1-P, FPA2014-52823-C2-2-P and the program Severo Ochoa (SEV-2014-0398). It is also supported by the European Commission under the FP7/EURATOM contract 605203 and by the Academy of Finland under the Finnish Centre of Excellence Programme 2012–2017 (Project No. 213503, Nuclear and Accelerator-Based Physics Research at JYFL). A.K. also acknowledges the Academy of Finland grants No. 275389 and 284516. I.D. and M.M. acknowledge the support of the German Helmholtz Association via the Young Investigators Grant No. VH-NG 627 (LISA- Lifetime Spectroscopy for Astrophysics). W.G. acknowledges the support of the UK Science \& Technology Faculties Council (STFC) under grant No. ST/F012012/1 and the University of Valencia. 
\end{acknowledgments}

\bibliography{Bibliography.bib}

\end{document}